# BARRIER RF STACKING*

W. Chou[†], FNAL, Batavia, IL 60510, USA
A. Takagi, KEK, Oho 1-1. Tsukuba, Ibaraki 305-0801, Japan

*Abstract*

This paper introduces a new method for stacking beams in the longitudinal phase space. It uses RF barriers to confine and compress beams in an accelerator, provided that the machine momentum acceptance is a few times larger than the momentum spread of the injected beam. This is the case for the Fermilab Main Injector. A barrier RF system employing Finemet cores and high-voltage solid-state switches is under construction. The goal is to double the number of protons per cycle on the production target for Run2 and NuMI experiments.

## 1 INTRODUCTION

The Fermilab Tevatron collider experiment, nicknamed Run2, is the most important high-energy physics program in the world at this moment. And it will be so for the next several years until the LHC at CERN starts operating, which is scheduled around 2007. The key parameter of a successful Run2 is the total integrated luminosity. The goal is 10-15 fb$^{-1}$. In addition to Run2, there is also a neutrino program, NuMI, at Fermilab. It uses the 120-GeV proton beams from the Main Injector (MI) to generate high intensity neutrino beams for a long baseline experiment at Soudan, Minnesota. This experiment will start in early 2005.

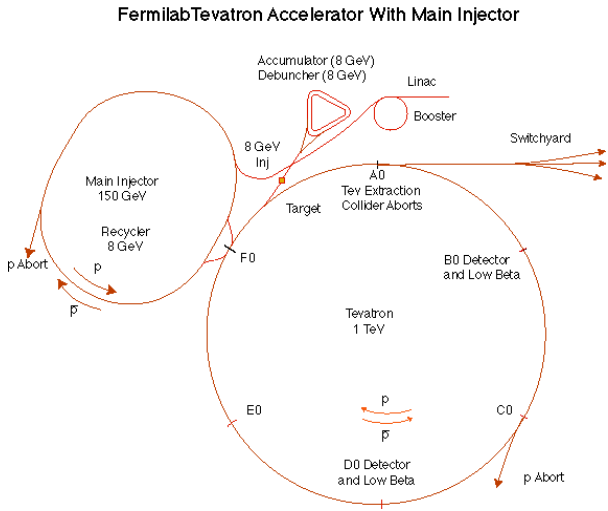

Figure 1: Fermilab accelerator complex.

In order to reach the goals of the luminosity in Run2 and the neutrino flux in NuMI, one needs to increase the proton intensity on the antiproton production target (Run2) and pion production target (NuMI), respectively. In the present Fermilab accelerator complex, as shown in Figure 1, the Booster is a bottleneck that limits the proton intensity on the targets. The number of protons per cycle from the Booster cannot exceed $5 \times 10^{12}$. Otherwise the beam loss would become prohibitive, as shown in Fig. 2.

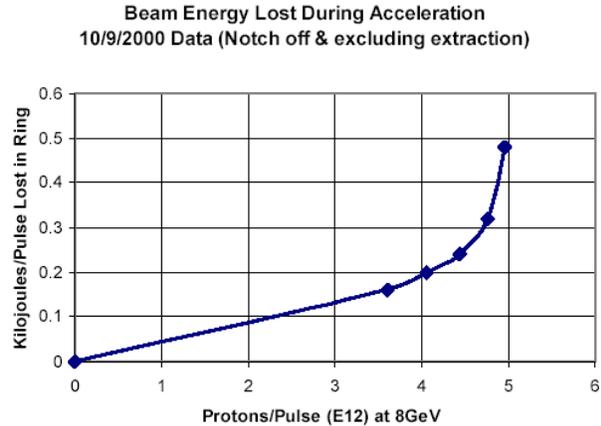

Figure 2. Fermilab Booster beam loss as a function of beam intensity. (Courtesy R. Webber)

To get around this bottleneck, one method is to use stacking. Namely, to put more than one Booster bunches into an Main Injector RF bucket. This is possible because the longitudinal acceptance of the Main Injector (0.4 eV-s) is larger than the longitudinal emittance of the Booster beam (0.1 eV-s). There are several possible ways to perform stacking. This paper will give a brief discussion of a method based on employing barrier RF systems. The goal is to double the number of protons per bunch in the Main Injector, which would then give twice as many protons on the production targets per cycle. The average production rate of antiprotons and neutrinos would increase 50-60%. For more information about this study the readers are referred to Ref. [1].

## 2 THE METHOD

A straightforward way to do barrier RF stacking is as follows. Inject two Booster batches into the MI, confine them by RF barriers, and then move the barriers to compress the beam. When the beam size is reduced to half of its original length (i.e., to the size of one Booster batch), the main RF system (53 MHz) in the MI is turned on to capture the beam and starts acceleration. The drawback of this approach is that the compression must be slow (adiabatic) in order to avoid emittance growth. This would lengthen the injection process and thus reduce the number of protons on the targets per unit time.

A better way, which was first proposed by J. Griffin [2], works as follows. Inject the Booster beams into the MI with a small energy offset (a few tens of MeV). Two RF barrier systems are employed. One is stationary, another moving. The stationary barrier serves as a firewall preventing particles from penetrating. The moving barrier bents the beam of successive injections so that the total beam length is continuously compressed. A detailed analysis and simulations have been performed by K-Y. Ng and can be found in Ref. [3].

There is a difference between the barrier RF stacking for Run2 and that for NuMI. In Run2, only a proton pulse of the length of one Booster batch is allowed to bombard the target, because the Antiproton Accumulator has the same circumference as the Booster. Therefore, the stacking process is 2-to-1, that is, two Booster batches compressed to the size of one. In NuMI, on the other hand, the proton pulse length is only limited by the MI circumference, which is equal to seven Booster batches. Thus, the proton pulse length can be as long as six Booster batches (leaving a gap for the extraction kicker rise-time). Correspondingly, the stacking process is 12-to-6, that is, twelve Booster batches compressed to the size of six. Figs. 3 and 4 illustrate the two different stacking processes.

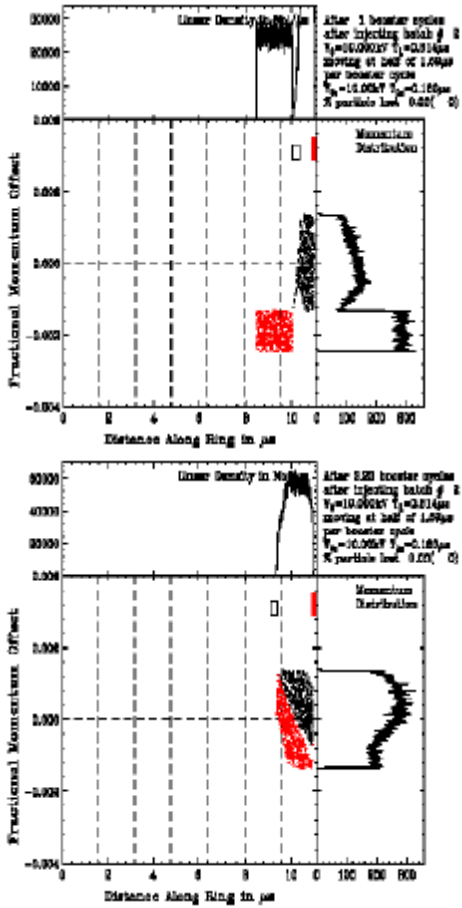

Figure 3. Barrier RF stacking for Run2. Top - The first batch (black) is confined and compressed by the RF barriers while the second batch (red) is injected in. Bottom - Two batches are confined and compressed to the length of one. The two small rectangles, one red one white, represent the two RF barriers. (Courtesy K-Y. Ng)

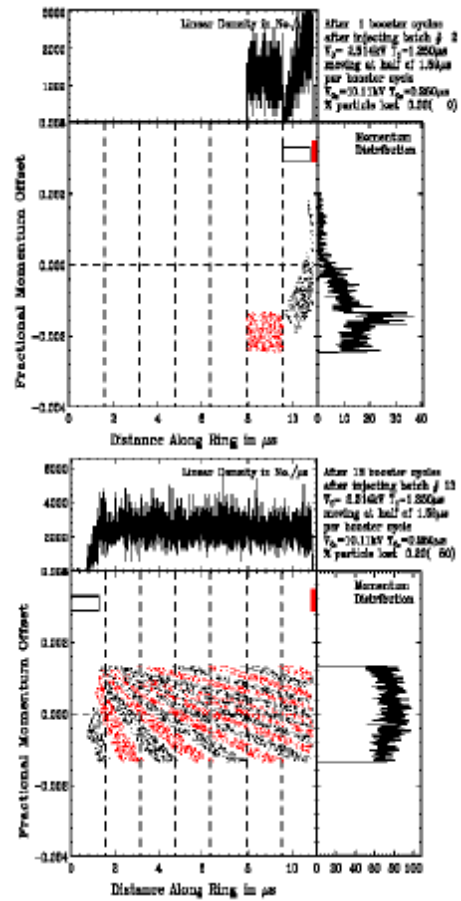

Figure 4. Barrier RF stacking for NuMI. Top - The first batch (black) is confined and compressed by the RF barriers while the second batch (red) is injected in. Bottom - Twelve batches are confined and compressed to the length of six. The two small rectangles, one red one white, represent the two RF barriers. (Courtesy K-Y. Ng)

## 3 TECHNICAL CONSIDERATIONS

### 3.1 Comparison with Slip Stacking

Stacking beams in the longitudinal phase space can also be achieved by the so-called slip stacking method. It uses two RF systems with slightly different frequencies to capture two successively injected bunches. When the two bunches are lined up in the same longitudinal position, they are merged into a common RF bucket. This method was first proposed at CERN in 1979 for the PS. [4,5] A main problem of this method was beam loading. It caused large particle loss and emittance blowup at high intensity operations, and limited the beam intensity to $8 \times 10^{12}$ protons per cycle in the PS. This method is now being reinvestigated at the Fermilab Main Injector with an improved beam loading compensation system. [6]

As a comparison, the barrier RF stacking has smaller beam loading effects. This is because the peak beam current during stacking is lower due to a debunched beam. This is a main advantage of this new method.

Another "advantage" of the barrier RF stacking is that it has never been tried before. So we don't know much about it. By contrast, we already know how hard the slip stacking is. Therefore, it is justified to test this alternative.

*3.2 Emittance Dilution and Particle Loss*

The simulation assumes 0.1 eV-s for the incoming Booster bunch longitudinal emittance (95%). During the process, the Booster beam is debunched, compressed, stacked, rebunched and captured by the 53 MHz RF bucket. The final bunch has an emittance of 0.32 eV-s. So the blowup factor is 3.2, which is tolerable, because the MI acceptance is 0.4 eV-s. The particle loss in the simulation is negligible.

*3.3 A Key Issue*

In order to make the barrier RF stacking work, a key issue is to keep the energy spread $\Delta E$ of the injected beam small. For the Fermilab MI, simulation shows that $\Delta E$ of the Booster beam must be below ±6 MeV (the smaller the better) so that the beam will be contained in the RF bucket after stacking.

However, the present Booster beam has a $\Delta E$ of three times larger at an intensity of $5 \times 10^{12}$ due to coupled bunch instability. Several measures are being tested to reduce the energy spread: (a) a longitudinal feedback system, (2) RF frequency modulation to provide Landau damping, [7,8] and (3) bunch rotation prior to extraction.

## 4 SYSTEM DESIGN

In order to perform beam stacking in the MI, a Fermilab-KEK-Caltech team is designing and fabricating a barrier RF system. The hardware is primarily funded by a US-Japan collaboration program.

An ideal barrier RF system is a wideband system rather than a resonant one, although the latter has also been used for this purpose. [9,10] One can use a wideband amplifier driving a 50 Ω gap to generate the required isolated voltage pulses, as is done in the Fermilab Recycler. [11] But this is an expensive approach. Instead, we adopt the design using an inductive device with a low quality factor, which is driven by high voltage solid-state switches. In the following sections, we give a brief description of this system.

*4.1 System Description and Parameters*

The system consists of an RF cavity and a power supply made of high voltage fast switches. It generates isolated square voltage pulses of both polarities. There are two different types of RF barriers:

- Stationary barrier: This is a series of bipolar pulses (+ and −) generated once every MI turn (11.2 μs), as shown in Fig. 5(a). This system is similar to the one that was built and tested by a Fermilab-KEK-HIMAC team for a different purpose, i.e., as an RF chopper. [12]
- Moving barrier: This is a series of separated bipolar pulses shown in Fig. 5(b). The spacing between +V and −V pulses varies from 0 up to 11 μs. They are also generated once every MI turn.

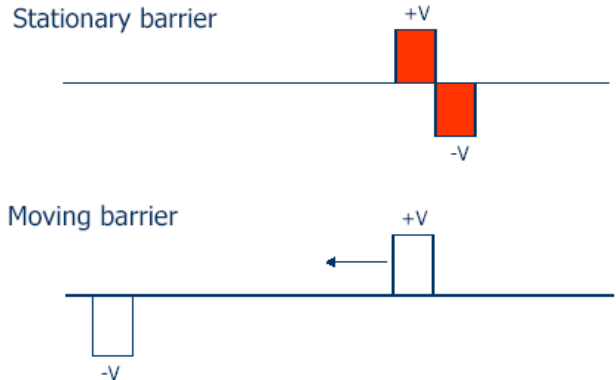

Figure 5. (a) Top: a stationary RF barrier consisting of a pair of bipolar pulses. Each pair is spaced by an MI revolution period. (b) Bottom: a moving RF barrier consisting of two isolated pulses with a variable gap in between. This pattern is repeated in every MI revolution period.

This barrier RF system works in burst mode. In other words, it generates a burst of pulses for a short period (about 200 ms), and then stops. The time between two bursts is fairly long (about 2 sec). Therefore, the duty factor of this system is low. This makes the use of solid-state switches possible.

There are some differences in the parameters of a RF stacking system for Run2 and that for NuMI. The following parameters in Table 1 were chosen as a first pass. It can be used for Run2 stacking. (NuMI stacking would require longer burst length.)

Table 1: Barrier RF System Parameters

| Pulse peak voltage | ±6 kV |
| --- | --- |
| Pulse maximum length | 0.6 μs |
| Pulse gap | 0 - 11 μs |
| Max pulse repetition rate | 1 MHz |
| Burst length | 200 ms |
| Burst repetition rate | 0.5 Hz |

*4.2 Cavity*

The cavity is inductive loaded and uses Finemet cores. Finemet is a magnetic alloy recently developed by the Hitachi Metals Ltd. in Japan. It has a nanocrystal structure (i.e., crystals of nanometer size). Compared with ferrite, it has considerably higher permeability in the frequency range of several MHz. It can also stand a much higher magnetic field at these frequencies. Its *Q* value is less than one. So it has a wide bandwidth. Fig. 6 shows a picture of a Finemet core. The specifications of the Finemet cores are listed in Table 2. A total of five cores will be used in this cavity.

___________________________________________

*Work supported by Universities Research Association, Inc. under contract No. DE-AC02-76CH03000 with the U.S. Department of Energy.
†chou@fnal.gov

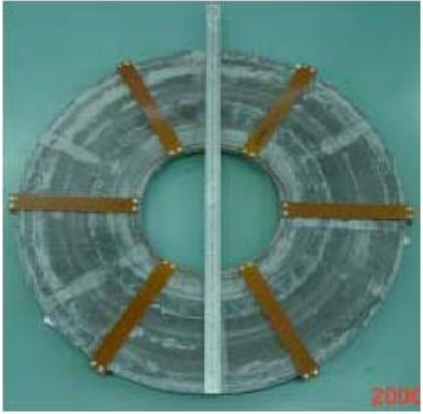

Figure 6. A Finemet core.

Table 2: Finemet Core Specifications

| Core outside diameter (OD) | 500 mm |
|---|---|
| Core inside diameter (ID) | 139.8 mm |
| Core width | 25 mm |
| Stainless steel mandrel OD | 139.8 mm |
| Stainless steel mandrel ID | 133 mm |
| Stainless steel mandrel thickness | 3.4 mm |
| Inductance per core (at 1 MHz) | 56 μH |
| Resistance per core (at 1 MHz) | 190 Ω |
| Quality factor (at 1 MHz) | 0.54 |

## *4.3 High-Voltage Fast Switch Circuit*

The switches need to have high peak voltage and high peak current. Because the load is inductive, the switches must be bipolar in order to avoid flyback when the pulse is terminated. The HTS 161-06-GSM solid-state switches made by Behlke Co. in Germany were chosen. The specifications are listed in Table 3.

Table 3: HTS 161-06-GSM Switch Specifications

| Maximum peak voltage | $2 \times \pm 8$ kV |
|---|---|
| Maximum peak current | $2 \times 60$ A |
| Maximum burst frequency | 2 MHz |
| Rise and fall time | 20 ns |
| Minimum pulse width | 200 ns |
| Minimum pulse spacing | 400 ns |
| Internal resistance | $2 \times 40$ Ω |

Compared with the RF chopper design in Ref. [12], a new challenge to the barrier RF design is the creation of a gap between +V and −V pulses. In the induction linac world, a barrier RF is sometimes called an "ear pulse." It is considered to be straightforward to obtain such a monopolar pulse. However, this is not the case for a barrier RF. The main difference is the repetition rate. An induction linac works at a few tens of Hz. A barrier RF works at ~ 1 MHz. At a low repetition rate, the circuit discharge time constant *L/R* is short compared with the time between two "ears." So the cavity has enough time to unload its energy stored during the pulse. At a high repetition rate, however, the stored energy in the cavity must be taken out before the next pulse comes. So the circuit design is more complicated. A preliminary design is shown in Fig. 7. L1 and R1 represent the cavity inductance and resistance. The distributed capacitance of the cavity is assumed to be 50 pF. R4-C2 is a snubber circuit, L2-R3 a damper circuit. Fig. 8 shows the voltage and current simulated by the code SPICE. They meet the requirements.

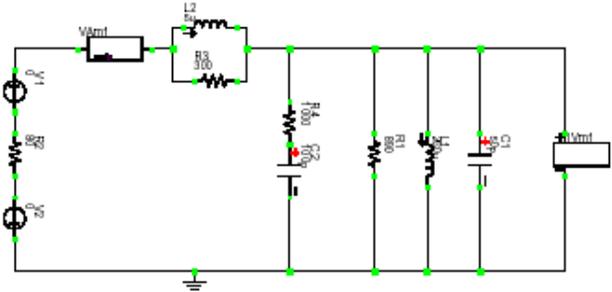

Figure 7. SPICE model of the circuit for the barrier RF power supply. L1-R1-C1 is a parallel representation of the cavity. R4-C2 in parallel to the cavity is a snubber. L2-R3 in series with the cavity is a damper. The internal resistance R2 of the two switches is also included.

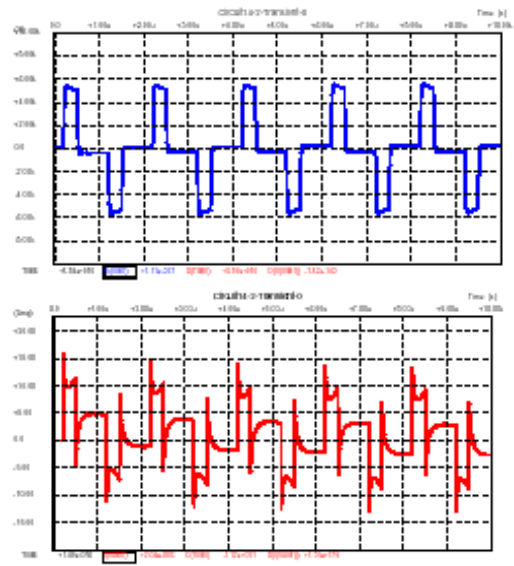

Figure 8. Voltage (top) and current (bottom) simulated by SPICE for the circuit in Fig. 7.

- 4 -

## 5 SUMMARY

A barrier RF system made of induction devices can be used for longitudinal stacking in an accelerator for increasing beam intensities. Compared with induction RF acceleration, barrier RF is much easier thanks to its low duty factor. An induction acceleration RF works in CW mode. A typical duty factor is 50%. A barrier RF, on the other hand, works in burst mode. Its duty factor is usually a fraction of a percent.

Finemet is an ideal material for barrier RF cavity cores. Its permeability is significantly higher than ferrite at a few MHz. A main concern, however, is its cost (a couple of hundreds US dollars per kilogram for a completed core purchased from Hitachi). The hope is that the price would go down as demands go up.

The HTS high-voltage fast switches meet the design requirements. A potential concern is the radiation shielding of the switch box. Because it must be placed next to the cavity in the tunnel in order to minimize stray inductance and capacitance.

A Fermilab-KEK-Caltech team is designing and building a barrier RF system that will be employed for stacking proton beams in the Main Injector. This system will be ready for test during the first half of 2003.

## 6 ACKNOWLEDGEMENT

This paper is based on a series of meetings at Fermilab focused on the study of barrier RF stacking. Many people made important contributions at different stages of this study, in particular, J. Griffin, K-Y. Ng, J. MacLachlan, D. Wildman, C. Bhat, I. Kourbanis, C. Ankenbrandt, M. Popovic and K. Koba. The authors are indebted to them.

A Caltech team led by D. Michael has been actively involved in this study.

The hardware is funded by the high intensity proton R&D program in a US-Japan collaboration in high-energy physics.